\documentclass[printer]{aa}  
\usepackage{natbib}
\bibpunct{(}{)}{;}{a}{}{,} 

\usepackage{graphicx}
\usepackage{txfonts}
\usepackage{amstext}

\begin{document}

   \title{Slowly rotating Bose--Einstein Condensate confronted with the rotation curves of 12 dwarf galaxies}

   \author{E. Kun\inst{1}, Z. Keresztes\inst{2}, L. \'{A}. Gergely\inst{3}}

   \institute{Department of Experimental Physics, University of Szeged, D\'om t\'er 9, H-6720 Szeged, Hungary\\
              \email{kun.emma0608@gmail.com}
                   \and
             Department of Theoretical Physics, University of Szeged, Tisza Lajos krt 84-86, H-6720 Szeged, Hungary\\
             \email{zkeresztes.zk@gmail.com}
                                \and
             Institute of Physics, University of Szeged, D\'om t\'er 9, H-6720 Szeged, Hungary\\
             \email{laszlo.a.gergely@gmail.com}
               }
\titlerunning{A test of the slowly rotating BEC model}


 
  \abstract
  {We assemble a database of $12$ dwarf galaxies, for which optical ($R$-band) and near-infrared ($3.6\mu m$) surface brightness density together with spectroscopic rotation curve data are available, in order to test the slowly rotating Bose--Einstein Condensate (BEC) dark matter model.}
  {We aim to establish the angular velocity range compatible with observations, bounded from above by the requirement of finite size halos, to check the modelfits with the dataset, and the universality of the BEC halo parameter $\mathcal{R}$.}
  {We construct the spatial luminosity density of the stellar component of the dwarf galaxies based on their $3.6\mu m$ and R-band surface brightness profiles, assuming an axisymmetric baryonic mass distribution with arbitrary axis ratio. We build up the gaseous component of the mass by employing a truncated disk model. We fit a baryonic plus dark matter combined model, parametrized by the $M/L$ ratios of the baryonic components and parameters of the slowly rotating BEC (the central density $\rho_\mathrm{c}$, size of the BEC halo $\mathcal{R}$ in the static limit, angular velocity $\omega$) to the rotation curve data.}
  {The $3.6\mu m$ surface brightness of six galaxies indicates the presence of a bulge and a disk component. The shape of the $3.6\mu m$ and $R$-band spatial mass density profiles being similar is consistent with the stellar mass of the galaxies emerging wavelength-independent. The slowly rotating BEC model fits the rotation curve of $11$ galaxies out of 12 within $1\sigma$ significance level, with the average of $\mathcal{R}$ as $7.51$ kpc and standard deviation of $2.96$ kpc. This represents an improvement over the static BEC model fits, also discussed. For the well-fitting $11$ galaxies the angular velocities allowing for a finite size slowly rotating BEC halo are less then $2.2\times 10^{-16}$ $s^{-1}$. For a scattering length of the BEC particle of $a\approx 10^6$ fm, as allowed by terrestrial laboratory experiments, the mass of the BEC particle is slightly better constrained than in the static case as $m\in[1.26\times10^{-17}\div3.08\times10^{-17}]$(eV/c$^{2}$).}
   {}


   \maketitle
%
%

\section{Introduction}

The pioneering work by Vera Rubin and her collaborators on optical (H$\alpha$) galaxy rotation curves proved the presence of an unknown form of matter \citep{Rubin1978,Rubin1985}. It was followed up by the radio (HI) observations, first systematically conducted by Albert Bosma \citep[e.g.][]{Bosma1977,Bosma1981}. Fritz Zwicky also concluded from the dynamic analysis of galaxy clusters the existence of some invisible material \citep{Zwicky1937}, referred as dark matter (DM).

Since then other evidence appeared for matter interacting only gravitationally, such as gravitational lensing \citep[e.g.][]{Wegg2016,Chudaykin2016}, or measurements on the cosmic microwave background radiation \citep{Planck2015}. Recent observations with the Planck satellite indicate that the DM makes up about one quarter of the energy of the Universe \citep{Planck2015,Planck2018}. 

Galactic astronomy cannot explain the observed rotation curves through luminous matter alone. Several DM-type mass density profiles were proposed to relax the problem of the missing mass. The Navarro-Frenk-White (NFW) DM model \citep{NFW1996} emerged from cold DM structure-formation simulations. The pseudo-isothermal halo model \citep{Gunn1972} has a core-like constant density profile avoiding the density singularity of the NFW model emerging at the center of the galaxies.

Supplementing other viable proposals, \citet{Bohmer2007} considered the possibility that DM could be in the form of a Bose--Einstein Condensate (BEC). They described DM as a non-relativistic, Newtonian gravitational BEC gas, obeying the Gross–Pitaevskii equation with density and pressure related through a barotropic equation of state. They fitted the Newtonian tangential velocity of the model with a sample of rotation curves of low surface brightness and dwarf galaxies, finding good agreement. 

\citet{Dwornik2015} tested the BEC DM model against rotation curve data of high and low surface brightness galaxies. Fits were of similar quality for the BEC and NFW DM models, except for the rotation curves exhibiting long flat regions, slightly better favouring the NFW profiles.

\citet{Kun2018mgrbec} confronted a non-relativistic BEC model of light bosons interacting gravitationally either through a Newtonian or a Yukawa potential with the observed rotational curves of 12 dwarf galaxies. The rotational curves of 5 galaxies were reproduced with high confidence level by the BEC model. Allowing for a small mass the gravitons resulted in similar performances of the fit. The upper mass limit for the graviton in this approach resulted in $10^{-26}$  eV {c}$^{-2}$.

\citet{Zhang2018} derived the tangential velocity of a test particle moving in a slowly rotating Bose--Einstein Condensate (srBEC)-type DM halo. In this paper we confront their model with the rotation curve of 12 dwarf galaxies. The rotational velocity is parametrized by the central density of the srBEC halo ($\rho_\mathrm{c}$), the radius of the static BEC halo ($\mathcal{R}$), and the angular velocity ($\omega$) of the srBEC halo. The value of $\mathcal{R}$ is determined by the scattering length $a$ and the mass $m$ of the DM particle. Therefore $\mathcal{R}$ is expected to be a universal constant and the different size of the srBEC halos should emerge due to the differences in their angular velocity.

In Section \ref{baryoniccomponent} we give the contribution of the baryonic component to the galaxy rotation curves. We present the model of the stellar component, we argue for a more sophisticated stellar model producing better results than the widely accepted exponential disk model, and we build up the $3.6 \mu m$ and $R$-band spatial luminosity density models to compare them to each other. At the end of this Section we present the model of the gaseous component. In Section \ref{section:darkmatter} we introduce the srBEC model. We address the maximum rotation of the srBEC halos, a novel concept advanced in relation with this model. In Section \ref{sec:besftmodels} we present and discuss the best-fit rotation curve models of 12 dwarf galaxies. In Section \ref{sec:discussion} we summarize our results and give final remarks.

\section{Baryonic model}
\label{baryoniccomponent}

\subsection{Stellar component}

The stellar contribution to rotational curves is derived based on the distribution of the luminous matter, deduced from the surface brightness of the galaxies. We follow \cite{Tempel2006} to derive the surface brightness density model, assuming the spatial luminosity density distribution of each visible component given by
\begin{equation}
l(a)=l(0)\exp\left[ -\left( \frac{a}{ka_\mathrm{0}}\right)^{{1/N}} \right].
\label{eq:spatlumdistr}
\end{equation}
Here $l(0)=hL {(4\pi q a_\mathrm{0}^3)}^{-1}$ is the central density, where $a_\mathrm{0}$ characterizes the harmonic mean radius of the respective component, and $k$ and $h$ are scaling parameters. Furthermore, $a=\sqrt{r^2+z^2 q^{-2}}$, where $q$ is the axis ratio, and $r$ and $z$ are cylindrical coordinates. From the measurements the projection of $l(a)$ onto the plane of the sky perpendicular to the line of sight, the surface luminosity is derived cf. \citet{Kun2017}:
\begin{equation}
S(R)=2 \sum_i^n q_i \int_R^\infty \frac{l_\mathrm{i}(a) a}{\sqrt{a^2-R^2}}da.
\label{eq:sr}
\end{equation}
Here $S(R)$ arises as a sum for $n$ visible components, and we assumed constant axis ratios $q_\mathrm{i}$. Equation (\ref{eq:sr}) was fitted to the observed surface luminosity profiles, assuming a constant axis ratio $q$. In the two-component stellar model the spatial mass density is 
\begin{equation}
\rho(a)= \Upsilon_\mathrm{b} l_\mathrm{b}(a)+\Upsilon_\mathrm{d} l_d(a),
\label{eq:rhoa}
\end{equation}
where $l_\mathrm{b}(a)$ and $l_\mathrm{d}(a)$ are the spatial luminosity densities of the bulge and disk components, and $\Upsilon_\mathrm{b}$ and $\Upsilon_\mathrm{d}$ are the respective mass-to-light ($M/L$) ratios (given in solar units).

It follows from the Poisson equation that for spheroidal shape matter, the rotational velocity squared in the galactic plane ($z=0$) induced by each stellar component is given by \citep{Tamm2005}:
\begin{equation}
v_{i,*}^2(R)=4 \pi q_\mathrm{i} G \int_0^R \frac{\rho_\mathrm{i}(r) r^2}{(R^2-e_\mathrm{i}^2 r^2)^{1/2}} dr,
\end{equation}
where $i={b,d}$, $G$ is the gravitational constant, $e_i=(1-q_\mathrm{i}^2)^{1/2}$ is the eccentricity of the $i$th stellar component, and $\rho_\mathrm{i}(r)$ is its mass density. 

\subsection{Exponential disk and Tempel--Tenjes models}

{\begin{table*}
\centering
\caption{Best-fit parameters describing the luminosity density distribution of the baryonic matter of dwarf galaxies at $3.6 \mu m$ (indicated by the superscript $^{NIR}$) and optical wavelengths (indicated by the superscript $^R$). The total luminosity of the galaxies ($L_\mathrm{b}$ for the bulge and $L_\mathrm{d}$ for the disk) is also presented.}
\label{table:gx_photometry}
\begin{tabular}{lcccccccccccc}
\hline
\hline
ID & $l(0)_b$ & $ka_{0,b}$ & $N_b$ & $q_b$ & $l(0)_\mathrm{d}$ & $ka_{0,d}$ & $N_\mathrm{d}$ & $q_\mathrm{d}$ & $L_\mathrm{b}$ & $L_\mathrm{d}$\\
(UGC)  & $\left(\frac{L_\odot}{kpc^3}\right)$ & ($kpc$) &  &  &  $\left(\frac{L_\odot}{kpc^3}\right)$  & ($kpc$) &  & &($10^9 L_\odot$) & ($10^9 L_\odot$)\\
  \hline
1281$^{\mathrm{NIR}}$ & -& -& -& -& $6.328\cdot10^8$& 0.872& 1.112& 0.156 & -& 2.548\\
1281$^{\mathrm{R}}$ & -& -& -& -& $3.056\cdot10^8$& 0.89& 1.075& 0.166 & - & 1.200\\
\hline
4325$^{\mathrm{NIR}}$ &$6.733\cdot10^8$& 0.046& 2.236& 0.769& $1.884\cdot10^8$& 2.930& 0.5& 0.110 & 0.594 &2.902\\
4325$^{\mathrm{R}}$  &$3.53\cdot10^7$& 0.612 & 0.986 & 0.808 & $9.531\cdot10^7$ & 2.290 & 0.758 & 0.095 & 0.156 &1.346\\
\hline
4499$^{\mathrm{NIR}}$ & -& -& -& -& $1.079\cdot10^{10}$& 0.085& 2.023& 0.095 & -& 2.161\\
4499$^{\mathrm{R}}$ & -& -& -& -&  $3.693\cdot10^9$& 0.114& 1.835& 0.099 &-& 0.659\\
\hline
5721$^{\mathrm{NIR}}$ &$6.650\cdot10^9$& 0.095& 1.31& 0.856& $6.23\cdot10^8$& 1.309& 0.401& 0.06& 0.442& 0.387\\
5721$^{\mathrm{R}}$ & $3.020\cdot10^9$ & 0.07 & 1.602 & 0.800 & $3.203\cdot10^8$ & 0.732& 0.701& 0.100 & 0.300 & 0.116\\
\hline
5986$^{\mathrm{NIR}}$ &$5.170\cdot10^9$& 0.065& 2.00& 0.81& $1.409\cdot10^9$& 3.011& 0.551& 0.100 & 3.468& 23.984\\
5986$^{\mathrm{R}}$ &$6.793\cdot10^8$ & 0.249 & 1.168 & 0.792 & $8.064\cdot10^8$ & 1.183 & 1.178 & 0.089 & 0.407& 6.070\\
\hline
6446$^{\mathrm{NIR}}$ & -& -& -& -& $1.286\cdot10^{10}$& 0.053& 2.217& 0.071 &-&1.432\\
6446$^{\mathrm{R}}$ & -& -& -& -& $3.837\cdot10^9$& 0.122 &1.975 &0.109& -& 1.991\\
\hline
7125$^{\mathrm{NIR}}$ &$9.499\cdot10^7$& 1.833& 0.7908& 0.6174& $5.041\cdot10^7$& 2.716& 1.388& 0.063 & 4.379 & 8.212\\
7125$^{\mathrm{R}}$ &$7.125\cdot10^7 $&  0.331 & 1.762 & 0.700 & $9.92\cdot10^7$ & 2.35& 1.134 & 0.109 & 1.492& 5.974\\
\hline
7151$^{\mathrm{NIR}}$ & -& -& -& -& $1.314\cdot10^{10}$& 0.374& 1.433& 0.078 &-& 8.538\\
7151$^{\mathrm{R}}$ & -& -& -& -& $2.341 \cdot 10^9$ & 0.359 & 1.504 & 0.076 &-& 1.840\\
\hline
7399$^{\mathrm{NIR}}$ &$1.674\cdot10^9$& 0.324& 1.096& 0.671& $3.763\cdot10^7$& 5.088& 0.144& 0.072 & 1.395 &1.324\\
7399$^{\mathrm{R}}$ &$1.505\cdot10^9$ & 0.071 & 1.714& 0.836& $1.053\cdot 10^8$ & 1.511 & 0.712 & 0.106 & 0.289& 0.367\\
\hline
7603$^{\mathrm{NIR}}$ & -& -& -& -& $9.362\cdot10^9$& 0.161& 1.479& 0.130& -& 1.006\\
7603$^{\mathrm{R}}$ & -& -& -& -&  $1.592\cdot10^9$ &0.539 & 0.891 & 0.112 & -& 0.472\\
\hline
8286$^{\mathrm{NIR}}$ &$2.789\cdot10^9$& 0.048& 2.063& 0.604& $1.286\cdot10^9$& 1.604& 0.801& 0.107 & 0.802& 7.114\\
8286$^{\mathrm{R}}$ &$6.601\cdot10^8$ & 0.058 & 2.321 & 0.689 & $6.953\cdot10^8$ & 1.117 & 0.845 & 0.147 & 1.739&2.061\\
\hline
8490$^{\mathrm{NIR}}$ & -& -& -& -& $3.080\cdot10^{10}$& 0.07& 1.88& 0.10 & -&1.640\\
8490$^{\mathrm{R}}$ & -& -& -& -&  $2.582\cdot10^9$ & 0.157 & 1.569 & 0.246& -&0.755\\
\hline
\end{tabular}
\end{table*}
\begin{table*}
\centering
\caption{Values of $\sigma$ and $\tau$ for the 12 dwarf galaxies of the sample, which give how much larger the $R$-band $M/L$ ratio of the bulge and disk are compared to those of the $3.6\mu m$ $M/L$ ratios.}
\label{table:sigmatau}
\begin{tabular}{ccccccccccccc}
\hline
\hline
 & 1281 & 4325 & 4499 & 5721 &  5986 & 6446 & 7125 & 7151 & 7309 & 7603 &8286 & 8490\\
 \hline
$\sigma$ &-& 3.808 & - & 1.473 & 8.521 &-&2.935&-&4.827&- & 0.461 & -\\
$\tau$ & 2.123 & 2.156 &3.279 & 3.336 & 3.951 & 0.719 &1.375 & 4.640 & 3.607 & 2.131 &3.452 &2.172\\
\hline
\end{tabular}
\end{table*}}
The SPARC database \citep{Lelli2016} offers robust mass models of a sample of 175 disk galaxies with Spitzer $3.6 \mu m$ photometry together with accurate rotation curves, well-suited to test rotation curve models. The largest number of dwarf galaxies were assumed to be bulgeless, and their photometry was fitted by an exponential disk model. The disk model is a widely explored in automatized modelling. We select 12 galaxies from this database, with the longest near-infrared (NIR) surface photometry profiles and accurate rotation curves, for which $R$-band counterparts are also available (this last criterion being motivated in the next subsection).

\begin{figure*}
\centering
\includegraphics[scale=0.4]{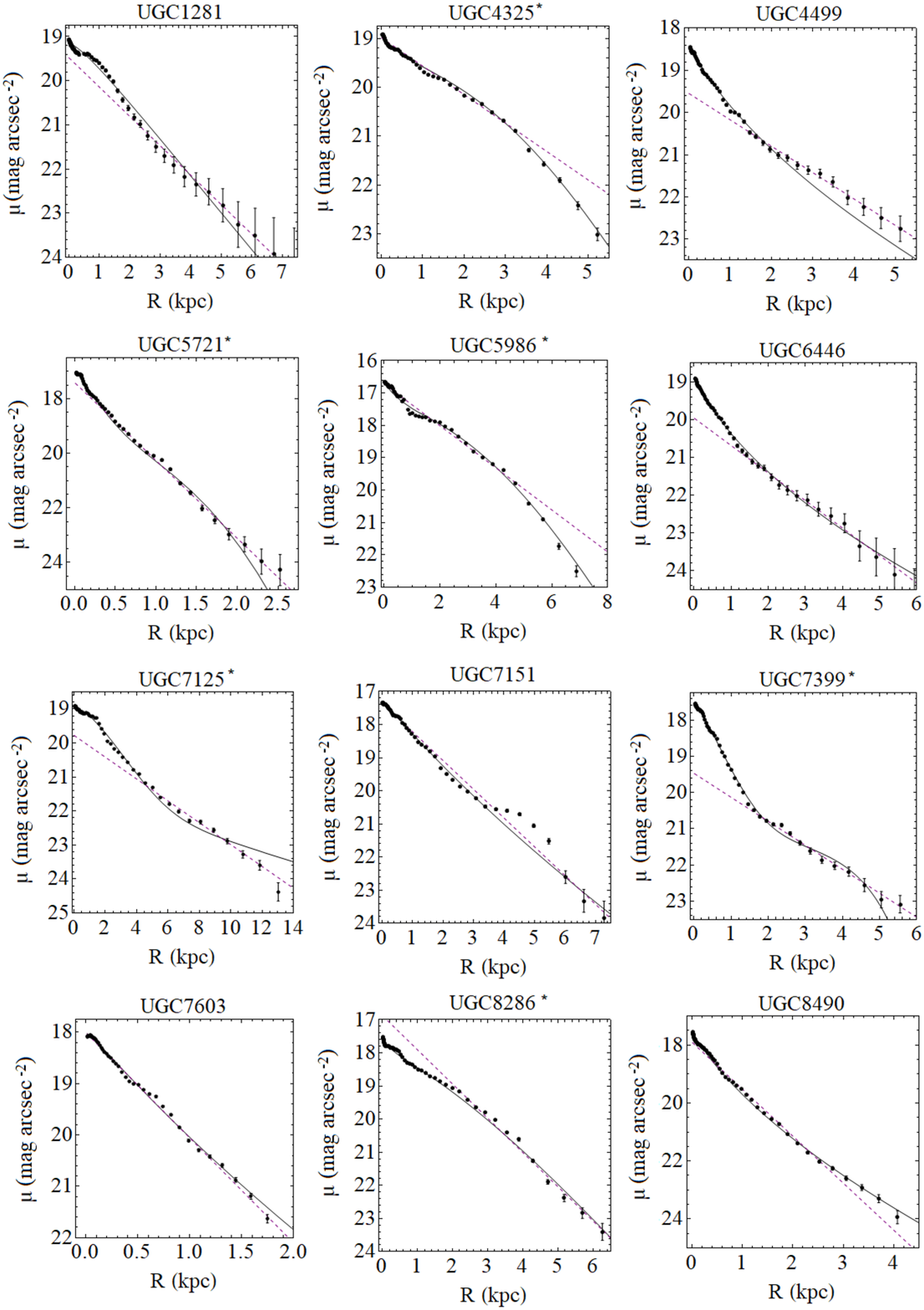}
\caption{Exponential disk model (purple dashed line) from the SPARC database and Tempel--Tenjes disk model in the present paper, or bulge+disk model (black continuous line) of the 12 dwarf galaxies. The $3.6 \mu m$ SPARC surface brightness data are presented by black dots with error-bars. The $^\star$ sign marks galaxies with two-component stellar model (bulge+disk).}
\label{fig:expttcaomp_dwarfs}
\end{figure*}

Unfortunately the exponential disk model in SPARC sometimes underestimates the luminosity of the inner region, in other cases under- or overestimates the outer region (see Fig. \ref{fig:expttcaomp_dwarfs}). 
Therefore we explored a more sophisticated Tempel--Tenjes model, moreover for half of the galaxies we allowed for both bulge and disk as indicated by their photometric data. We binned the NIR surface brightness profile of the galaxies on a logarithmic scale to smooth out possible small-scale inhomogenities in them. The best-fit baryonic parameters are presented in Table \ref{table:gx_photometry} (the respective galaxy names carry the superscript NIR). In Fig. \ref{fig:expttcaomp_dwarfs} we show the best-fit Tempel--Tenjes models for the chosen galaxies, along with their exponential disk fit from the SPARC. It is clear that the Tempel--Tenjes model has a better fit to the surface brightness data.

\subsection{Near-infrared and R-band spatial luminosity models}
\label{irrband}

We took the $R$-band (effective central wavelength $634.9$ nm, FWHM $106.56$ nm) surface brightness data of the same 12 late-type dwarf galaxies from the Westerbork HI survey of spiral and irregular galaxies, to build up their $R$-band photometric models \citep{Swaters1999,Swaters2002,Swaters2009}. These measurements were made with the 2.54 m Isaac Newton Telescope on La Palma in the Canary Islands. We again fitted the data with the Tempel--Tenjes model. Galaxies described by a two-component surface brightness model (bulge+disk) at $3.6\mu m$ are described by a two-component model in the $R$-band too. For the absolute $R$-magnitude of the Sun $\mathcal{M}_{\odot,R}=4.42^m$ \citep{Binney1998} was adopted. The best-fit parameters describing the $R$-band spatial luminosity density of these 12 dwarf galaxies are given in Table \ref{table:gx_photometry} (the galaxy names carrying the superscript $R$). Compared to the $3.6\mu m$ data, the $R$-band data result in lower luminosities for almost all of the galaxies (one exception is the galaxy UGC6446).

Earlier studies indicate that the near-infrared (NIR) $M/L$ ratio depends weakly on the color, several models predicting its constancy in the NIR over a broad range of galaxy masses and morphologies, for both the bulge and the disk \citep[e.g.][and references therein]{McGaugh2014}. The NIR surface photometry provides the most sensitive proxy to the stellar mass, as shown by e.g. \citet{Verheijen2001}. We build the spatial mass density distribution of the disk (and bulge) employing the NIR and $R$-band surface brightness models, and Eq. (\ref{eq:rhoa}).

The total mass of the stellar component should not depend on the wavelength at which the galaxies are observed. The masses of the bulge and the disk should be the same for the $3.6 \mu m$ and $R$-band measurements, i.e.:
\begin{eqnarray}
\Upsilon_\mathrm{NIR, b} L_\mathrm{NIR, b} (=M_\mathrm{NIR, b})=\Upsilon_\mathrm{R, b} L_\mathrm{R, b} (=M_\mathrm{R, b}),\\
\Upsilon_\mathrm{NIR, d} L_\mathrm{NIR, d} (=M_\mathrm{NIR, d})=\Upsilon_\mathrm{R, d} L_\mathrm{R, d} (=M_\mathrm{R, d}),
\end{eqnarray}  
where $\Upsilon$ is the $M/L$ ratio, $L$ is the total luminosity, $M$ is the total mass. We give the total luminosities in Table \ref{table:gx_photometry} based on the best-fit surface brightness models of the galaxies. Then the $R$-band $M/L$ ratios are:
\begin{eqnarray}
\Upsilon_\mathrm{R, b}=\frac{L_\mathrm{NIR, b}}{L_\mathrm{R, b}}\Upsilon_\mathrm{NIR, b}=\sigma \Upsilon_\mathrm{NIR, b},\\
\Upsilon_\mathrm{R, d}=\frac{L_\mathrm{NIR, d}}{L_\mathrm{R, d}}\Upsilon_\mathrm{NIR, d}=\tau \Upsilon_\mathrm{NIR, d},
\label{eq:mperlbulgediskr}
\end{eqnarray} 
for the bulge and disk, respectively. The values for $\sigma$ and $\tau$ are given in Table \ref{table:sigmatau}, calculated based on the $R$ to $NIR$ ratio of the total luminosities of the $12$ galaxies of the sample.

\begin{figure*}
\centering
\includegraphics[scale=0.42]{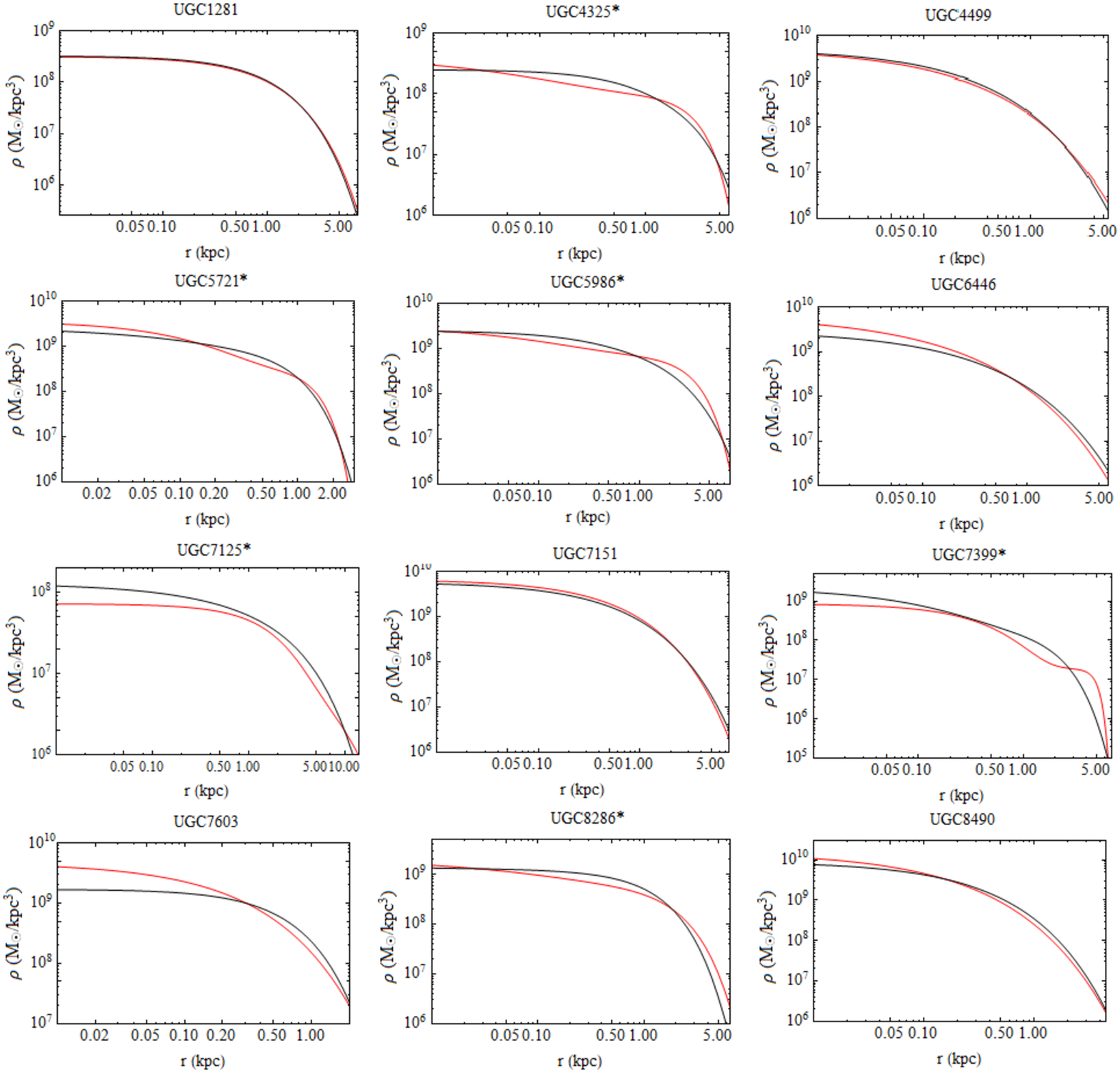}
\caption{Mass density models of the 12 galaxies at $3.6 \mu m$ (red line) and in $R$-band (black line). The coordinate $r$ is measured in the galactic plane (where $a=r$, because $z=0$). The $^\star$ sign marks galaxies with two-component stellar model (bulge + disk).}
\label{fig:rhos}
\end{figure*}
In Fig \ref{fig:rhos} we plot the mass densities of the 12 dwarf galaxies employing the best-fit surface brightness density models (from Table \ref{table:gx_photometry}) Stellar population models and earlier studies on the conversion between the NIR flux and stellar mass suggest that the typical value of the stellar $M/L$ in NIR should be at about $0.5 M_\odot/L_\odot$ \citep[e.g.][and references therein]{Eskew2012, McGaugh2014}. We assume $\Upsilon_\mathrm{NIR, d}\equiv0.5$ to derive the mass density from the luminosity density, and for those galaxies with bulge $\Upsilon_\mathrm{NIR, b}\equiv0.5$, in order to calculate through Eqs. (\ref{eq:mperlbulgediskr}) how much larger is the $M/L$ of the disk (and the bulge where it applies) than that of the NIR $M/L$s. In case of the galaxies of the present sample $\sigma$ (where it applies) and $\tau$ are given in Table \ref{table:sigmatau}. The predicted shape of the spatial mass densities is similar for both the $R$- and NIR-bands.

The SPARC $3.6\mu m$ photometry samples the surface brightness of the galaxies from a region five-ten times closer to the centre of the galaxies, out of the same region where the $R$-band observations end. Due to their good resolution, and the fact that they are the closest proxy to the stellar mass distribution, we employ the SPARC $3.6 \mu m$ data to model the stellar component of the baryonic mass of the galaxies in the next section, to test the slowly rotating BEC model.

\newpage
\subsection{Gaseous component}

Observations of galaxies show that for a large fraction of dwarf galaxies the rotation velocity of the gas (measured by emission lines) is close to the rotation velocity of the stellar component (measured by absorption lines) \citep[e.g.][]{Rhee2004,Lelli2016}. Therefore for these galaxies it is necessary to involve a gaseous contribution to the baryonic component of their rotation curves. For this purpose we include an additional velocity square of an exponential disk \citep{BT1987}:
\begin{equation}
v_\mathrm{gas}^2(R)= 4 \pi G \Sigma_{0} R_\mathrm{d} y^2 \left[I_0(y) K_0(y)-I_1(y) K_1(y)\right],
\end{equation}
where $\Sigma_{0}$ is the central surface mass density, $R_\mathrm{d}$ is the scale length of the disk, $y \equiv R/2R_\mathrm{d}$, and $I$ and $K$ are the modified Bessel functions. The mass of the disk within radius $R$ is 
\begin{equation}
M_\mathrm{d}(R)=2\pi \Sigma_0 R^2_\mathrm{d} \left[ 1 - \exp \left( -\frac{R}{R_\mathrm{d}} \right) \left( 1+\frac{R}{R_\mathrm{d}}\right)\right],
\end{equation}
while its total mass is:
\begin{equation}
M_\mathrm{tot,d}=2 \pi \Sigma_0 R^2_\mathrm{d}.
\end{equation}
We employ these equation in $R-R_\mathrm{t}$ by a introducing a truncation radius $R_\mathrm{t}$, where $R_\mathrm{t}$ denotes that radius outside of which the gaseous component is not negligible. To build up the contribution of the gaseous component to the baryonic rotation curves we fitted this truncated exponential disk model to the discrete values of the gas velocity given in the SPARC database. The best-fit parameters are given in Table \ref{table:vrot_bestfit}.

\section{Dark matter model}
\label{section:darkmatter}

\subsection{The slowly rotating BEC-type dark matter component}

The equatorial radius of the srBEC DM halo is given by \citet{Zhang2018}
\begin{equation}
R_0\left(\frac{\pi}{2}\right)= \frac{\pi}{k}\left(1+\frac{9}{4}\Omega ^2\right),
\label{eq:rnull}
\end{equation}
where
\begin{eqnarray}
\Omega ^2&=&\frac{\omega^2}{2\pi G\rho_\mathrm{c}}=0.02386\times \nonumber\\
&&\times \left(\frac{\omega }{10^{-16}\;{\rm s}^{-1}}\right)^2\times \left(\frac{\rho_\mathrm{c}}{10^{-24}\;{\rm g/cm^3}}\right)^{-1},
\end{eqnarray}
$\rho_\mathrm{c}$ is the central density and $\omega$ the angular velocity of the srBEC halo (assumed to be in rigid rotation). 
In the non-rotating case $\Omega =0$ and $R_0\left(\pi/2\right)=\mathcal{R}=\pi/k$, the radius of the static BEC DM halo $\mathcal{R}$, being determined by the mass $m$ and scattering length $a$ of the DM particle through
\begin{equation}
k=\sqrt{\frac{G m^3}{a \hbar ^2}},
\end{equation}
where $\hbar$ is the reduced Planck-constant. The tangential velocity squared $v_\mathrm{srBEC}^2$ of massive test particles rotating in the BEC galactic DM halo is given in the first order of approximation as in the equatorial plane of the galaxies by
\begin{eqnarray}
&&v_\mathrm{srBEC}^2(R)=\frac{4G\rho_\mathrm{c} \mathcal{R}^2}{\pi}\times \nonumber\\
&&\Bigg[\left(1-\Omega ^2\right)\frac{\sin \left(\pi R/\mathcal{R}\right)}{\pi R/\mathcal{R}}-\left(1-\Omega ^2\right)\cos \frac{\pi R}{\mathcal{R}} +\frac{\Omega ^2}{3}\left(\frac{\pi R}{\mathcal{R}}\right)^2\Bigg], \nonumber\\
\end{eqnarray}
or equivalently\footnote{This equation corrects Eq.~(103) of \cite{Zhang2018}. When they substituted the definition of $\Omega ^2$ from their Eq.~(52) to Eq.~(103) missed the term $\left(\rho_\mathrm{c}/10^{-24}\;{\rm g/cm^3}\right)^{-1}$ from the right-side of their Eq.~(52) due to a misprint. We thank T. Harko for pointing this out to us.},
\begin{eqnarray}
\hspace{-1.5cm}&&v_\mathrm{srBEC}^2\left( \;{\rm km^2/s^2}\right)= 80.861\times \left(\frac{\rho_\mathrm{c}}{10^{-24}\;{\rm g/cm^3}}\right)\times \left(\frac{\mathcal{R}}{{\rm kpc}}\right)^2\times \nonumber\\
\hspace{-1.5cm}&&\Bigg[\left(1-\Omega ^2\right)\left[\frac{\sin \left(\pi R/\mathcal{R}\right)}{\pi R/\mathcal{R}}-\cos \frac{\pi R}{\mathcal{R}}\right] +\frac{\Omega ^2}{3}\left(\frac{\pi R}{\mathcal{R}}\right)^2\Bigg]. \nonumber\\
\label{eq:becv2}
\end{eqnarray}

{\begin{figure*}
\centering
\includegraphics[scale=0.22]{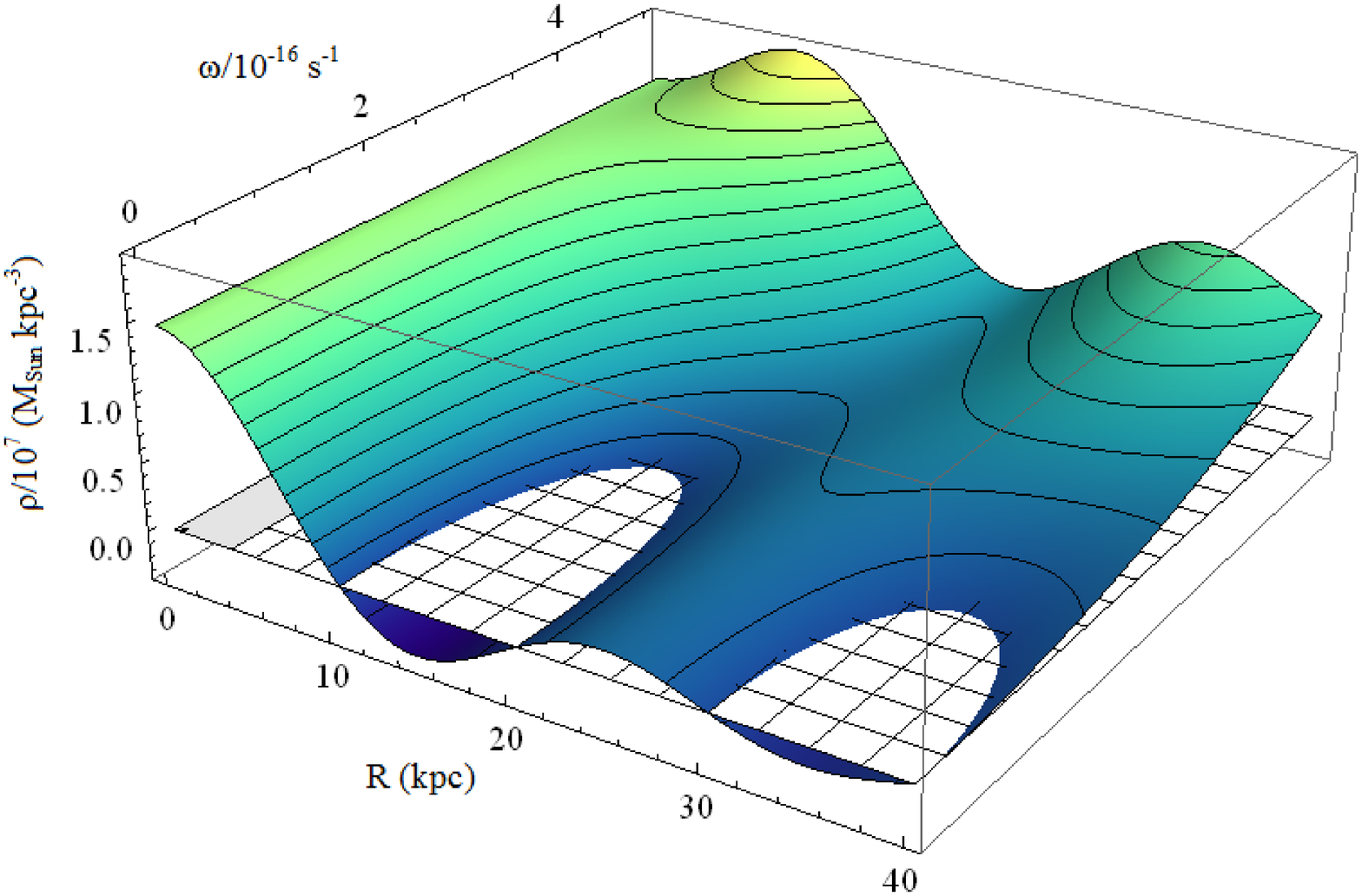}
\includegraphics[scale=0.22]{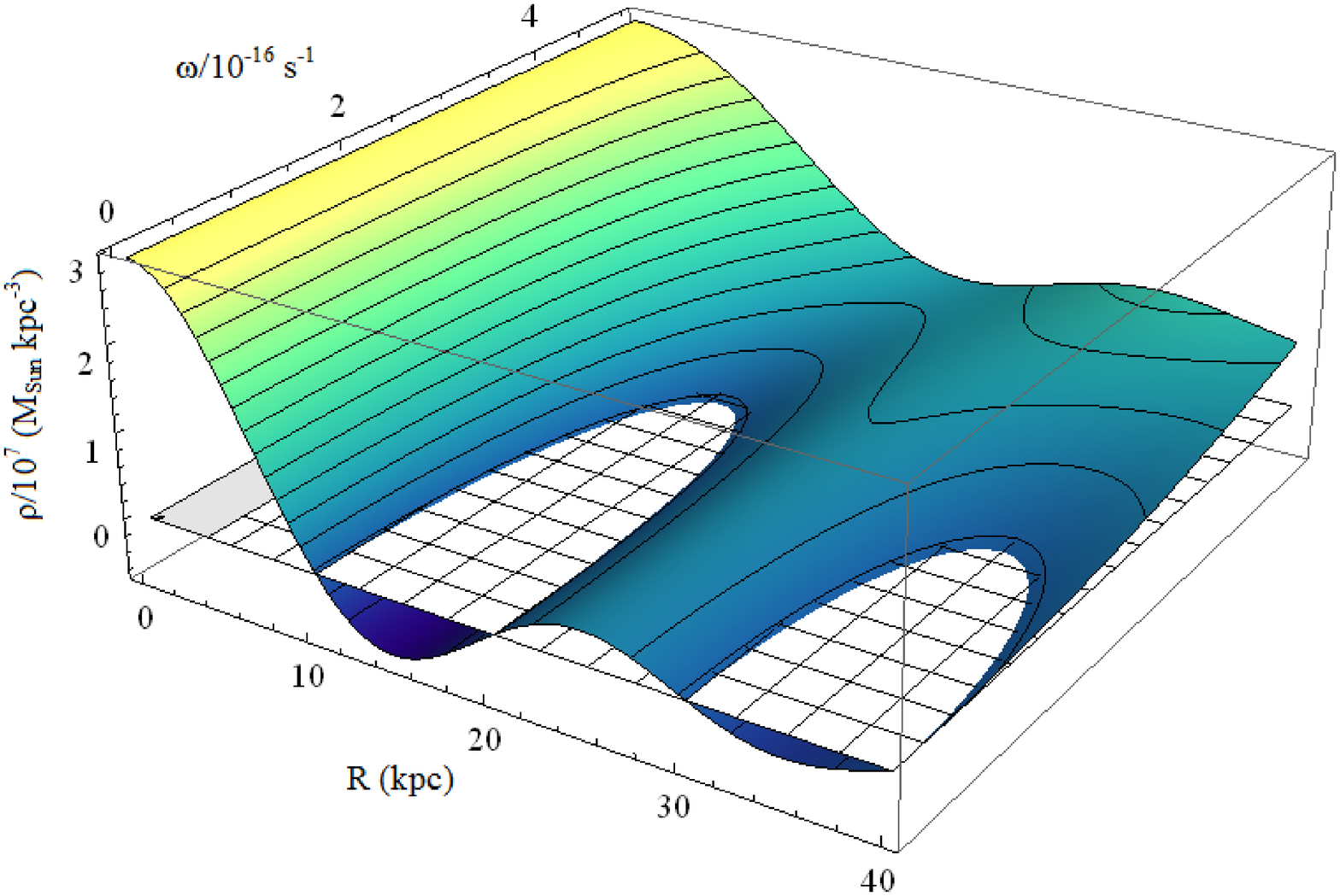}
\includegraphics[scale=0.22]{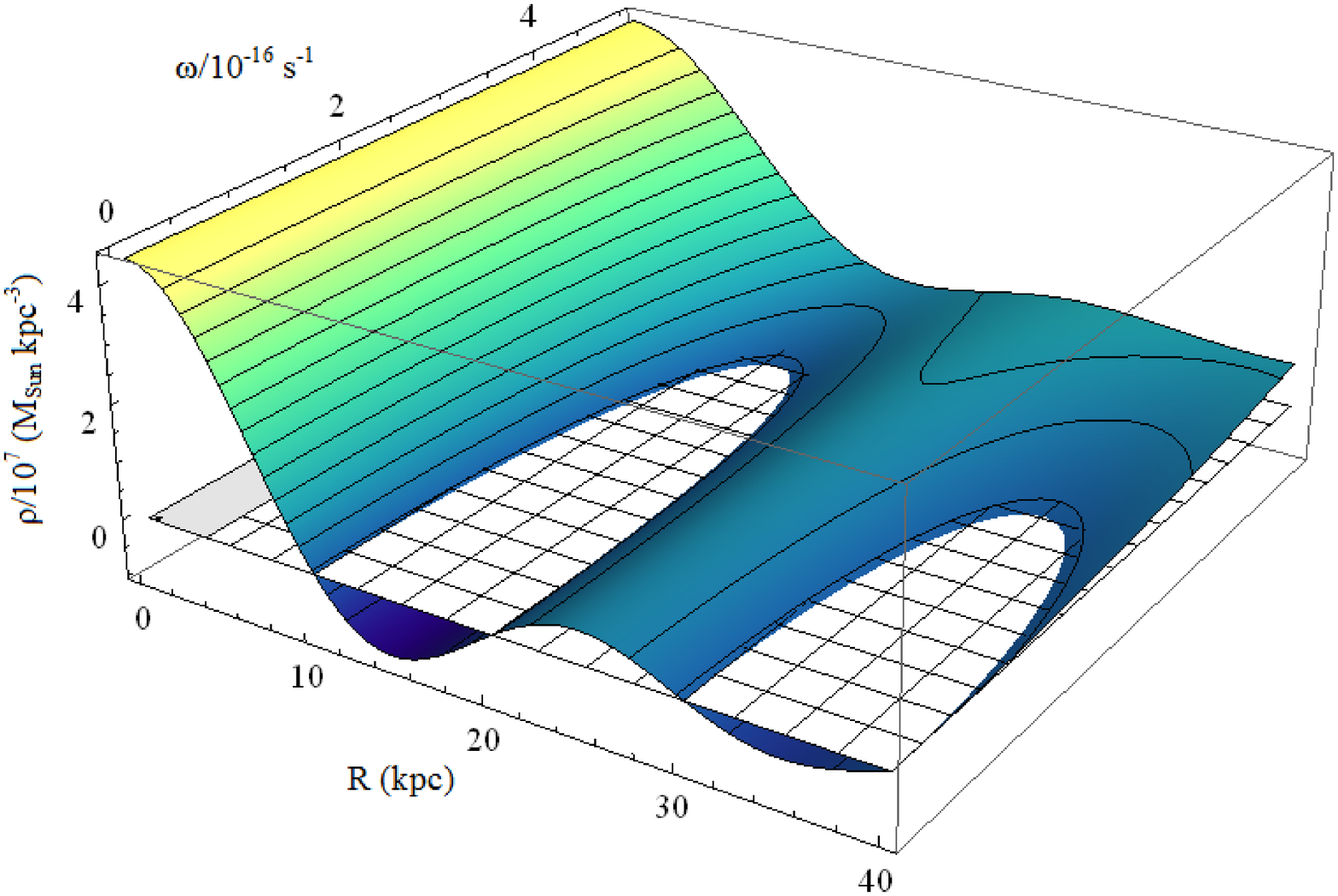}\newline
\caption{Density of the srBEC halo (coloured surface), as a function of the distance measured from the rotation axis of the galaxy in its equatorial plane ($R$, on the $x$-axis), and of the angular velocity ($\omega$, on the $y$-axis). The $\rho=0$ level surface is also indicated. Model parameters are: the size of the BEC halo in the static limit $\mathcal{R}=10$ kpc, the central density is $\rho_\mathrm{c} =1 \times 10^{-24} g/cm^3$ (left panel), $\rho_\mathrm{c} =2 \times 10^{-24} g/cm^3$ (middle), and $\rho_\mathrm{c} =3 \times 10^{-24} g/cm^3$ (right). With increasing BEC DM halo rotation, its density does not drop to zero, rather it exhibits a positive density extending to infinity. We consider the density profile realistic only when and until it first reaches the zero-level. The fastest rotation velocity $\omega$ of a realistic srBEC halo (that can have zero density at a given radius) depends on the central density $\rho_\mathrm{c}$, larger $\rho_\mathrm{c}$ resulting in higher limiting $\omega$.}
\label{fig:multirho1}
\end{figure*}
\begin{figure*}
\centering
\includegraphics[scale=0.25]{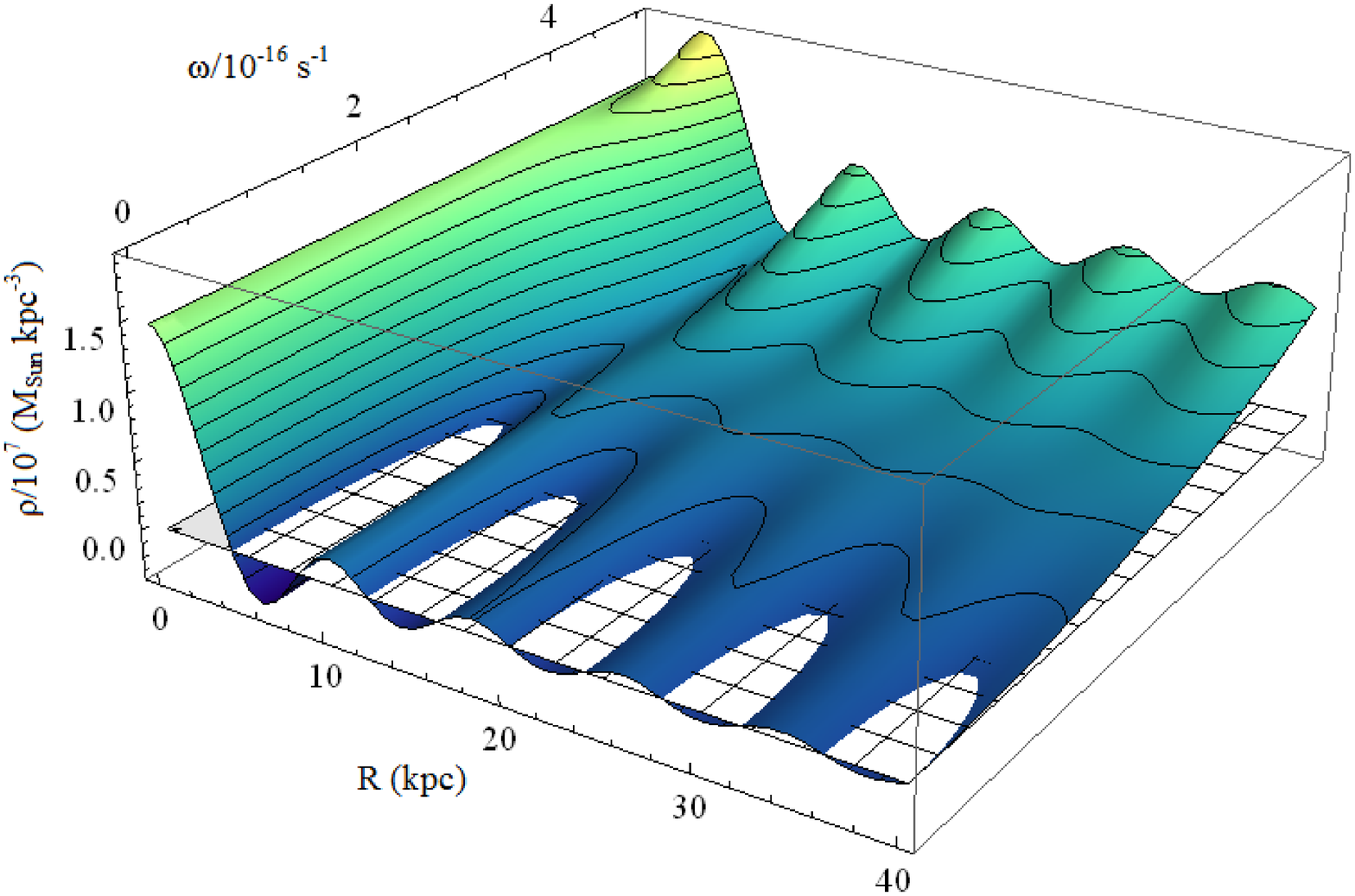}
\includegraphics[scale=0.33]{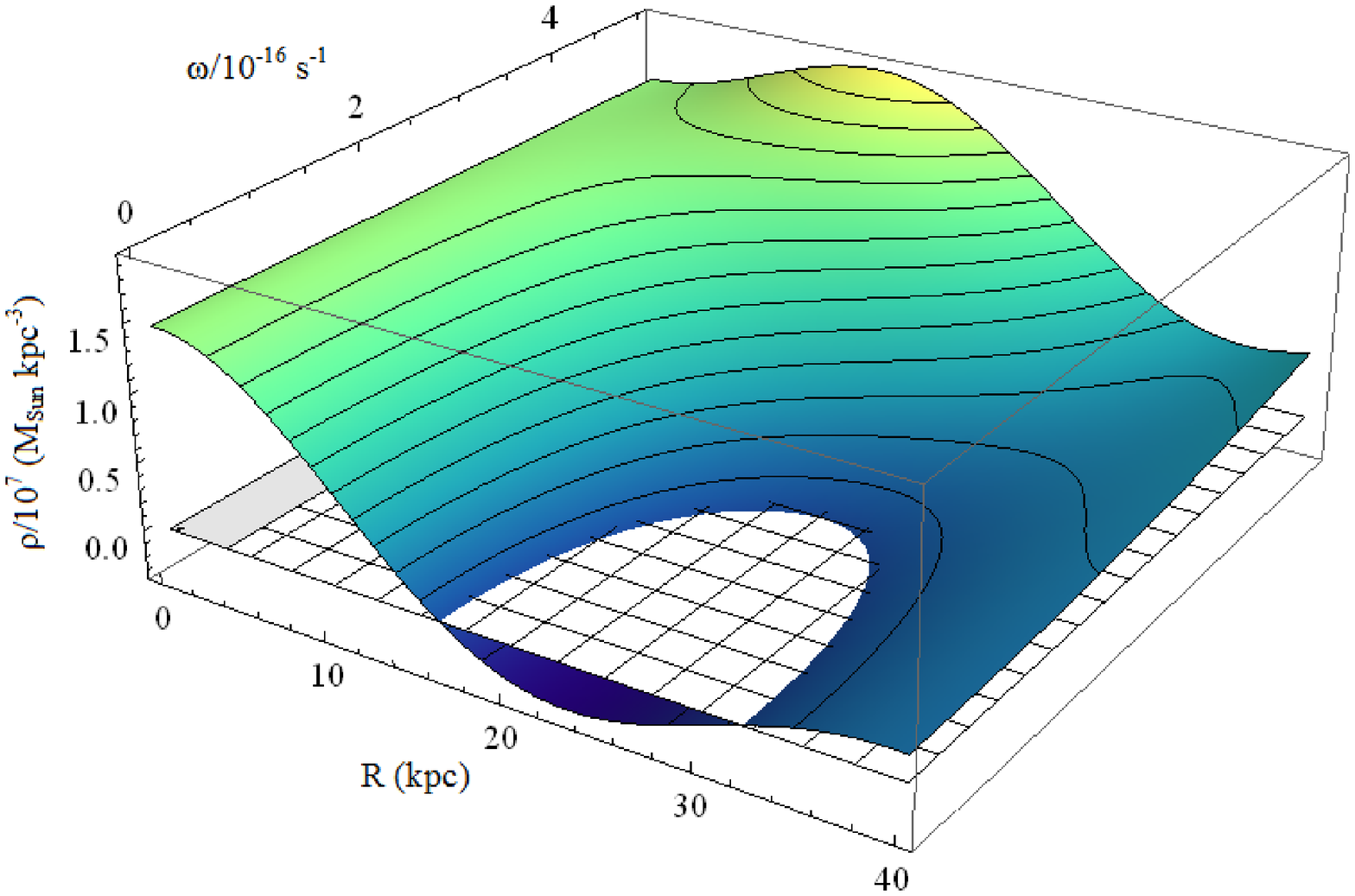}\newline
\caption{Density profile of the srBEC model (coloured surface), as a function of the distance measured from the rotation axis of the galaxy  in its equatorial plane ($R$, on the $x$-axis), and of the angular velocity ($\omega$, on the $y$-axis). For comparison we also indicate the $\rho=0$ level. Three model parameters are the follows. The central density is $\rho_\mathrm{c} =1 \times 10^{-24} g/cm^3$, and the size of the BEC halo in static limit is $\mathcal{R}=4$ kpc for the left-side panel and $\mathcal{R}=16$ kpc for the right-side panel. On the left panel one can see that for small angular velocity values the srBEC density reaches zero, after which a cut-off of the model has to be applied in order to have only positive densities. This procedure also ensures the finite size of the srBEC halo. For larger angular velocities however the density oscillates around a positive value, hence no cutoff is possible and the srBEC halo extends to infinity. Comparing the left- and right-sided panels it seems the fastest angular velocity $\omega$ of a finite srBEC halo does not depend on $\mathcal{R}$.}
\label{fig:multirho2}
\end{figure*}}

\subsection{On the maximal rotation of the slowly rotating BEC halo}
\label{maxspeed}

When formulating the srBEC model, \cite{Zhang2018} applied first order corrections to the density and radius of the DM halo. On Fig. \ref{fig:multirho1} we present the density profile of the srBEC halo as a function of the distance from the center of the galaxy measured in its equatorial plane and of the angular velocity of the DM halo, for three values of the central density $\rho_\mathrm{c}$.

For fast rotation the halo density although oscillating, is positive at all radii, meaning that the halo size is infinite. For slower rotation however, at some finite radius the density reaches zero, where the model should have a cut-off (otherwise it is continued through negative densities). The two regimes are separated by a limiting omega value, the fastest angular velocity allowing for a finite srBEC halo (having zero density at a given radius). This limiting $\omega$ increases together with the value of the central density $\rho_\mathrm{c}$.

On Fig. \ref{fig:multirho2} we present again the density profile of the srBEC halo, varying this time the size $\mathcal{R}$ of the static BEC halo. The highest $\omega$ giving a finite size halo does not seem to depend on the size of the static BEC halo $\mathcal{R}$, only on the central density $\rho_\mathrm{c}$. While for small $\omega$ the trigonometric term in Eq. (\ref{eq:becv2}) dominates, for larger $\omega$ the monotonic $r^2$ term is dominant. In this paper we consider only finite-size srBEC models, thus those possessing an upper limit for $\omega$.

\section{Rotation curve model of 12 dwarf galaxies}
\label{sec:besftmodels}

\begin{center}
\begin{table*}
\centering
\begin{tabular}{l|cccc|ccc|cccccc}
\hline
\hline
ID &$\Sigma_{0}$ & $R_\mathrm{d}$ & $R_\mathrm{t}$ & $M_\mathrm{tot,g}$ &$\Upsilon_\mathrm{b}$ &$\Upsilon_\mathrm{d}$& $M_\mathrm{tot,s}$  & $\rho_\mathrm{c}$ & $\mathcal{R}$ &  $\omega$ &  $M_\mathrm{srBEC}$ & $\chi^2$ & $1\sigma$\\
\hline
(UGC)  & $\left(10^7 \frac{M_\odot}{kpc^2}\right)$ & $\left(kpc\right)$ & $\left(kpc\right)$ & $10^9 M_\odot$ & $\left(\frac{M_\odot}{L _\odot}\right)$ & $\left(\frac{M_\odot}{L _\odot}\right)$ &$10^9 M_\odot$ & $\left(10^{-24} \frac{g}{cm^3}\right)$ & $\left(kpc\right)$ & $\left(10^{-16} \frac{1}{s}\right)$ &  $10^9 M_\odot$ &  &\\
\hline
1281 & 2.34 & 1.54 & 1.35 & 0.35 & - & 0.10 & 0.25 & 1.216 & 4.992 & 1.232 & 3.02 & 1.29 & 24.58 \\
4325 & 2.41 & 5.35 & 0.60 & 4.35 & 4.12& 1.34& 6.34 & 0.77& 5.0& 1.65& 2.15 & 3.55 & 4.71 \\
4499 & 2.78 & 3.41 & 1.46 & 2.03 & - & 0.52& 1.13 & 0.731& 7.409& 1.560& 6.57 & 5.39  &7.03 \\
5721 & 2.22 & 2.43 & 0 & 0.82 & 0.59& 2.63& 1.28 & 1.213& 6.532& 0.676& 6.46 & 7.73  & 21.35 \\
5986 & 1.12 & 9.18 & 0 & 5.93 & 0.09 & 0.22 & 5.59 & 1.447  &7.884 & 2.193 & 15.67 &1.25  &12.64 \\
6446 & 1.83 & 9.17 & 1.06 & 9.67 &- & 2.01& 2.88 & 0.655& 8.413& 1.476& 8.62 & 2.69  &15.93 \\
7125 & 1.07 & 22.29 & 0 & 33.4 & 0.38& 0.34& 4.49 & 0.105& 14.381& 0.702& 7.33 & 1.33  &10.42 \\
7151 & 3.21 & 2.34 & 1.57 & 1.10 & - & 0.26& 2.23 & 0.745  &6.128  &2.046 & 4.19 &6.46  &9.30 \\
7399 & 1.92 & 3.20 & 0.44 & 1.24 & 1.34& 0.1& 2.00 & 2.68& 6.0& 1.89& 11.62 & 6.69  &7.03 \\
7603 & 1.26 & 2.53 & 0 & 0.51 &- & 0.37 & 0.38 & 2.692  &3.793 & 0.978 & 2.80 & 8.54  &10.42 \\
8286 & 1.42 & 5.48 & 1.00 & 2.68 & 0.56& 0.47& 3.76 &  0.410& 11.540& 0.063& 11.79 & 7.26  &14.81\\
8490$^\star$ &2.15 & 2.59 & 0.91 & 0.29 &- & 1.25& 2.06 & 0.80& 8.071& 1.64& 9.31 & 39.84  &29.93\\
\hline
\end{tabular}
\caption{Best-fit parameters of the rotational curve models of 12 dwarf galaxies. The best-fit central surface mass density ($\Sigma_{0}$), the scale length ($R_\mathrm{d}$), and the truncation radius ($R_\mathrm{t}$) of the gaseous component can be found in columns 2-4. Best-fit $M/L$ ratio ($\Upsilon_\mathrm{b}$) for the bulge (where applicable) and $M/L$ ratio ($\Upsilon_\mathrm{d}$) for disk are presented in the 6th and 7th columns. The best-fit parameters of the slowly rotating BEC model are given in columns 9-11: the central density of the rotating BEC halo ($\rho_\mathrm{c}$), size of the static BEC halo ($\mathcal{R}$), and the angular velocity of the rotating BEC halo ($\omega$). The $\chi^2$ of the fits and the $1\sigma$ significance levels are also presented. The only galaxy that cannot be fitted within $1\sigma$ is marked by $^\star$. The total masses of the gaseous ($M_\mathrm{tot,g}$), the stellar ($M_\mathrm{tot,s}$), and the slowly rotating BEC components ($M_\mathrm{srBEC}$) are also given in columns 5, 8, 12, respectively.}
\label{table:vrot_bestfit}
\end{table*}
\end{center}

In the previous sections we gave the contribution of the baryonic sector (Section \ref{baryoniccomponent}.) and the slowly rotating BEC-type DM halo (Section \ref{section:darkmatter}.) to the combined rotation curve models. Then the model rotation curve in the equatorial pane of the galaxy is given as \citep{Rodrigues2018}
\begin{equation}
v_\mathrm{rot}=\sqrt{v_\mathrm{gas}|v_\mathrm{gas}|+\Upsilon_\mathrm{b} v_\mathrm{b}|v_\mathrm{b}|+ \Upsilon_\mathrm{d} v_\mathrm{d}|v_\mathrm{d}|+v^2_\mathrm{srBEC}},
\label{eq:vrot}
\end{equation}
where $v_\mathrm{gas}$, $v_\mathrm{b}$, $v_\mathrm{d}$ and $v_\mathrm{srBEC}$ are the contributions of the gaseous component, the bulge (where it applies), the disk, and the DM halo to the rotation curves.

 When fitting Eq. (\ref{eq:vrot}) to the observed rotation curves, we apply a non-linear least-squares method to perform the fit with error$^{-2}$ weights, minimizing the residual sum of squares ($\chi^2$) between the data and the model. We are interested in such models, where the mass density of the halo drops to zero for a given radius, therefore we set an upper limit for $\omega$, such that we allow only fits which results in finite size halos (see Section \ref{maxspeed}). This limit is dynamically changing during the fit with $\rho_\mathrm{c}$. The fitted parameters are $\Upsilon_\mathrm{b}$ and $\Upsilon_\mathrm{d}$ for the stellar component, $\rho_\mathrm{c}$, $\mathcal{R}$ and $\omega$ for the srBEC component.  Fitting the $M/L$ ratios we are able to reveal the maximal performance of the srBEC model.
 
\begin{figure*}
\centering
\includegraphics[scale=0.33]{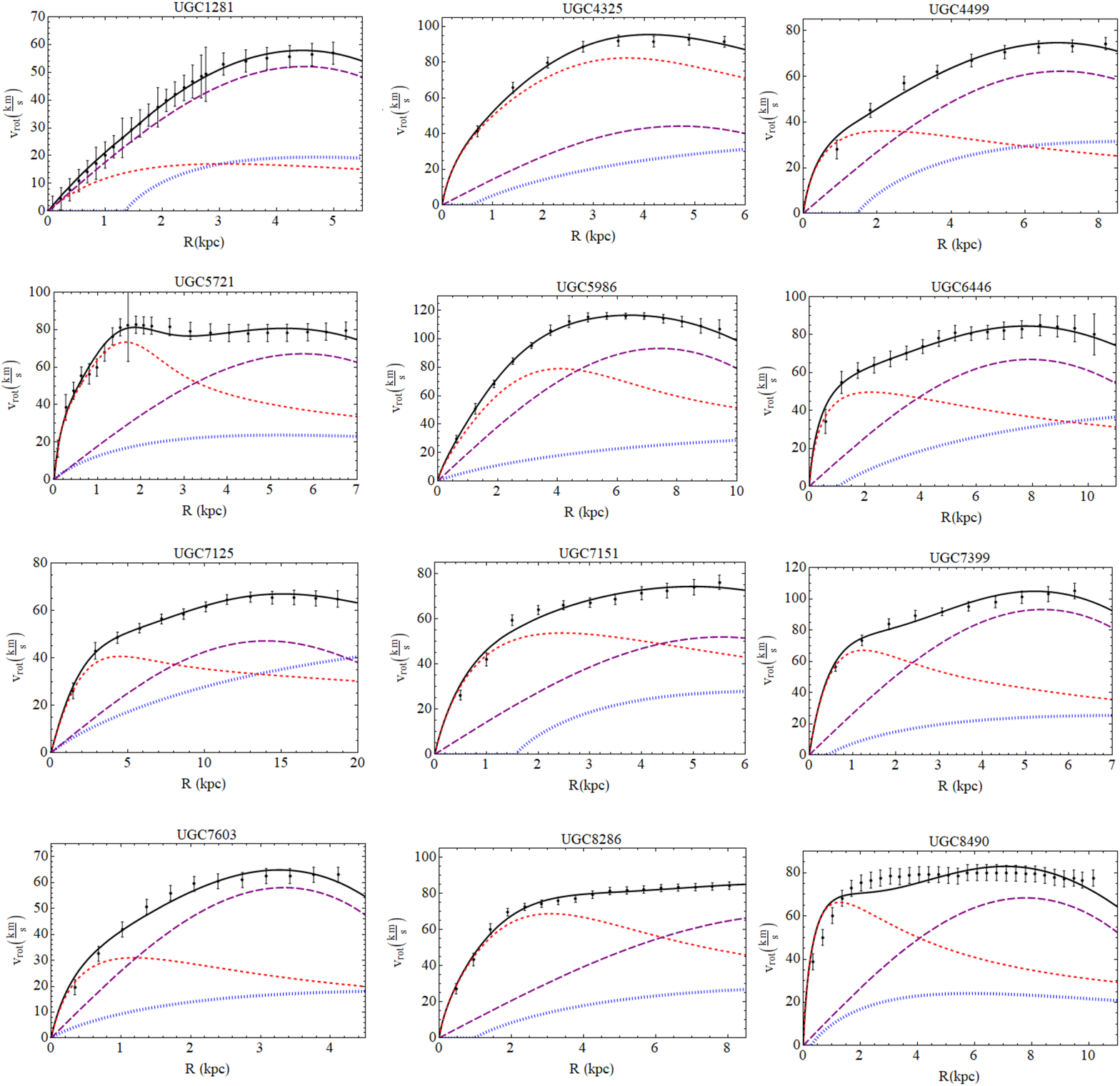}
\caption{Best-fit rotational curves of the dwarf galaxy sample. The dots with error-bars denote the observed rotational velocity curves. The fitted model, composed by a baryonic and a srBEC component is represented by the black curve. The red short-dashed curve draws the contribution of the stellar component, the blue tiny-dashed curve shows the contribution of the gas, and the long-dashed purple curve denotes the contribution of the srBEC-type DM halo to the rotation curve--models.}
\label{fig:vrot_dwarfs}
\end{figure*}

  The parameters of the best-fit galactic rotation curves, composed by a baryonic and a srBEC-type DM component are presented in Table \ref{table:vrot_bestfit}, and the best-fit rotation curves are shown in Fig.~\ref{fig:vrot_dwarfs} along the observed ones. The combined model fits the dataset within the $1\sigma$ confidence level in case of 11 dwarf galaxies out of 12. 
 
The size of the static BEC halo $\mathcal{R}$ is expected to be uniform for all of the galaxies, as it only depends on the mass and scattering length of the particle forming the BEC halo. From our fitting-procedure the average value of $\mathcal{R}$ emerged as 7.51 kpc, with standard deviation 2.96 kpc (see Table \ref{table:vrot_bestfit}).

In Section \ref{irrband} we derived the stellar mass density from the NIR luminosity density of the galaxies assuming $M/L$ ratios equal to $0.5$ in order to calculate how much larger are the R-band $M/L$ ratios compared to the NIR ones, and to plot the NIR and R-band mass density curves (the total mass of the stellar component should not depend on the observational band). By fitting the $M/L$ ratios (together with the srBEC parameters) to the rotational curve data we got different $M/L$s. Hence these galaxies may hold diverse stellar populations resulting in different luminosity characteristics \citep[e.g.][]{Bell2001,Bell2003}.

\section{Summary and final remarks}
\label{sec:discussion}

 In this paper we assembled photometric data and rotation curves of 12 late-type dwarf galaxies in order to test the srBEC DM--model from the SPARC database ($3.6 \mu m$ photometry) and the Westerbork HI survey of spiral and irregular galaxies ($R$-band photometry). Our particular interests were in 1) establishing the limiting angular velocity below which the model leads to finite size halos, 2) how well the model fits the dataset and 3) whether one of its parameters, the size of the BEC halo $\mathcal{R}$ in the static limit is really universal. 

\begin{table}
\centering
\begin{tabular}{l|cccccc}
\hline
\hline
ID  &$\Upsilon_\mathrm{b}'$ &$\Upsilon_\mathrm{d}'$& $\rho_\mathrm{c}'$ & $\mathcal{R'}$ & $\chi^2$ &$1\sigma$\\
\hline
(UGC) & $\left(\frac{M_\odot}{L _\odot}\right)$ & $\left(\frac{M_\odot}{L _\odot}\right)$ &$\left(10^{-24} \frac{g}{cm^3}\right)$ & $\left(kpc\right)$ & &\\
\hline
1281 & - & 0.11 & 1.201 & 5.011 & 1.50 &25.66\\
4325$^\star$ & 4.87 & 1.38 & 0.476 & 5.253 & 6.42 &5.89\\
4499 & - & 0.57 & 0.679 & 8.094 & 6.52 &8.18\\
5721 & 0.59 & 2.63 & 1.214 & 6.574 & 7.86 &22.64\\
5986 & 0.09 & 0.23 & 1.359 & 8.579 & 1.52&13.74\\
6446 & - & 2.03 & 0.631 & 9.105 & 3.31 &17.03\\
7125 & 0.32 & 0.98 & 0.059 & 15.36 & 1.56 &11.54\\
7151 & - & 0.27 & 0.714 & 6.806 & 7.62 &10.42\\
7399 & 1.34 & 0.03 & 2.623 & 6.221 & 7.27 &8.18\\
7603 & - & 0.36 & 2.763 & 3.794 & 10.10 &11.54\\
8286 & 0.56 & 0.47 & 0.411 & 11.54 & 7.26 &15.94\\
8490$^\star$ & - & 1.28 & 0.753 & 8.843 & 43.28 &31.00\\
\hline
\end{tabular}
\caption{Best-fit parameters of the rotational curve models of 12 dwarf galaxies. The rotational curves are composed of baryonic matter and a non-rotating BEC component. The fitted parameters are the M/L of the bulge ($\Upsilon_\mathrm{b}'$, where it applies) and the disk ($\Upsilon_\mathrm{d}'$), the central density of the non-rotating BEC halo ($\rho_\mathrm{c}'$) and size of the static BEC halo $\mathcal{R'}$. The two galaxies that cannot be fitted within $1\sigma$ are marked by $^\star$.}
\label{table:vrot_bestfit_nonrotbec}
\end{table}
We investigated whether the widely employed exponential disk model accurately describes the surface brightness of the galaxies, and found necessary to employ a more complicated model than the exponential one to correctly estimate the luminosity of the inner region of these galaxies. We built up the $3.6\mu m$ and $R$-band spatial luminosity densities of the galaxies fitting the Tempel--Tenjes model to their surface brightness densities. For six galaxies a two-component model (bulge+disk) described their surface brightness density more accurately then the disk model. We found the near infrared luminosity of almost all galaxies larger compared to the $R$-band one, leading to higher $M/L$ ratios in $R$-band in order to generate the same stellar mass. We added a gaseous component by fitting a truncated exponential disk to the gas velocity given in the SPARC database.

The stellar component+gas+slowly rotating BEC combined rotation curve model fits the dataset within the $1\sigma$ confidence level in case of 11 dwarf galaxies out of 12. The size of the static BEC halo $\mathcal{R}$, related to the BEC particle characteristics, hence expected to be the same for all of galaxies, has an average value of $\bar{\mathcal{R}}=7.51$ kpc, with standard deviation as $2.96$ kpc (see Table \ref{table:vrot_bestfit}). The best-fit limiting angular velocity which allows for a finite size slowly rotating BEC halo is $<2.2\times 10^{-16}$ $s^{-1}$ for the well-fitting 11 galaxies. Its average value is $1.32\times 10^{-16}$ $s^{-1}$, with standard deviation $0.66\times 10^{-16}$ $s^{-1}$. Based on the total masses (Table \ref{table:vrot_bestfit}) the slowly rotating BEC-type DM dominates the rotation curves of 9 galaxies out of 12 (exceptions are UGC 4325, UGC 6446, UGC 7125).

The mass $m$ of the BEC particle depends on its scattering length $a$ and the size of the static BEC halo $\mathcal{R}$ \citep{Bohmer2007}:
\begin{equation}
m=6.73\times 10^{-2} [a\mathrm{(fm)}]^{1/3} [\mathcal{R}\mathrm{(kpc)}]^{-2/3} \mathrm{eV}.
\end{equation}
Terrestrial laboratory experiments render the value of $a$ to be $\approx 10^6$ fm \citep[e.g][]{Bohmer2007}. Hence the mass of the BEC particle falls into the range $m\in[1.26\times10^{-17}\div3.08\times10^{-17}]$(eV/c$^2$) based on the best-fits of the srBEC model to the rotation curves of the present galaxy sample. The lower limit is given by galaxy UGC7125 having the largest static BEC halo ($\mathcal{R}=14.381$ kpc), and the upper limit from UGC7603 having the smallest one ($\mathcal{R}=3.793$ kpc). It is worth to note, that UGC7125 also has the smallest ($\rho_\mathrm{c}=0.105\times 10^{-24}$  g/cm$^3$), while UGC7603 the largest central density ($\rho_\mathrm{c}=2.692\times 10^{-24}$  g/cm$^3$) among these galaxies. A slightly different lower limit on $m$ emerges when assuming a static BEC model, the mass of the BEC particle falling into the range $m\in[1.21\times10^{-17}\div3.08\times10^{-17}]$(eV/c$^2$). Again, the two limits are constrained by the galaxies UGC7125 (from below, $\mathcal{R'}=15.36$ kpc) and UGC7603 (from above, $\mathcal{R'}=3.794$ kpc). We also note that UGC7125 possesses the longest, while UGC7603 the shortest observed rotation curve in the sample, hence the size of the static BEC halo seems to correlate with the length of the rotation curves.
 
Finally we discuss whether the slow rotation improves over the fits. By setting $\omega=0$ we fit a static BEC model to the rotational curve data, obtaining best-fit parameters given in Table \ref{table:vrot_bestfit_nonrotbec}. Comparison shows that the finite size srBEC model gave slightly better fits, which are below $1\sigma$ in 11 cases as compared to only 10 cases for the static BEC halo fits. In the static case the average value of $\mathcal{R'}$ emerged as $8.42$ kpc, with a standard deviation of $3.35$ kpc, as compared to the rotating BEC case with average of $7.51$ kpc and standard deviation of $2.96$ kpc. In a srBEC dark matter halo, the tangential velocity of a test particle is larger than in the static case at the same position \citep{Zhang2018}, also the plateau of the rotation curves is slightly lifted. With $\omega=0$ the fitting process favours larger $\mathcal{R'}$s to lift the plateau to give the same performance. This is why the average value of $\mathcal{R'}$ is larger than that of $\mathcal{R}$..

According to our rotation curve analysis, the srBEC halo with suitable constrained angular velocity values proves to be a viable DM model. However the steep decrease of either the static or the slowly rotating BEC rotation curves raises doubts on whether such a halo could be well fitted with galaxy lensing data.

\begin{acknowledgements}
We thank Tiberiu Harko and Maria Crăciun for suggesting to add the gas component to the baryonic sector. The authors acknowledge the support of the Hungarian National Research, Development and Innovation Office (NKFIH) in the form of the grant 123996. The work of Z. K. was supported by the János Bolyai Research Scholarship of the Hungarian Academy of Sciences and by the UNKP-18-4 New National Excellence Program of the Ministry of Human Capacities.
\end{acknowledgements}

\end{document}